\documentclass{aastex631}

\usepackage{amsfonts,amsmath,amssymb}

\graphicspath{{./}{figures/}}

\received{}
\revised{}
\accepted{}
\submitjournal{ApJ}

\shorttitle{Reionization History}
\shortauthors{Krishak et al.}

\begin{document}

\title{Gaussian Process Reconstruction of Reionization History}

\email{aditi16@iiserb.ac.in, dhiraj@imsc.res.in}

\author[0000-0002-8991-3070]{Aditi Krishak}
\affiliation{Department of Physics, Indian Institute of Science Education and Research Bhopal, Bhopal 462066, India}

\author[0000-0001-7041-4143]{Dhiraj Kumar Hazra}
\affiliation{The  Institute  of  Mathematical  Sciences,  HBNI,  CIT  Campus, Chennai  600113,  India}
\affiliation{INAF/OAS Bologna, Osservatorio di Astrofisica e Scienza dello Spazio,
		Area della ricerca CNR-INAF, via Gobetti 101, I-40129 Bologna, Italy}

\begin{abstract}
We reconstruct the history of reionization using Gaussian process regression. Using the UV luminosity data compilation from Hubble Frontiers Fields we reconstruct the redshift evolution of UV luminosity density and thereby the evolution of the source term in the ionization equation. This model-independent reconstruction rules out single power-law evolution of the luminosity density but supports the logarithmic double power-law parametrization. We obtain reionization history by integrating ionization equations with the reconstructed source term. Using optical depth constraint from Planck Cosmic Microwave Background observation, measurement of UV luminosity function integrated till truncation magnitude of -17 and -15, and derived ionization fraction from high redshift quasar, galaxies and gamma-ray burst observations, we constrain the history of reionization. In the conservative case we find the constraint on the optical depth as $\tau =0.052\pm0.001\pm0.002$ at 68\% and 95\% confidence intervals. We find the redshift duration between  10\% and 90\% ionization to be $2.05_{-0.21-0.30}^{+0.11+0.37}$. Longer duration of reionization is supported if UV luminosity density data with truncation magnitude of -15 is used in the joint analysis. Our results point out that even in a conservative reconstruction, a combination of cosmological and astrophysical observations can provide stringent constraints on the epoch of reionization.

\end{abstract}


\section{Introduction}\label{sec:intro}
In the late matter-dominated epoch of the Universe, the neutral hydrogen and helium atoms in the Inter-Galactic Medium (IGM) were ionized by the photons emitted from the sources formed during structure formation. Observations of quasar spectroscopy have confirmed that the IGM was ionized before $z\sim4$ with a neutral hydrogen fraction of about 1 in 10000 as we do not observe the~\cite{Gunn-Peterson} trough. Simulation of large-scale structure formation also reveals that the sources of high energetic photons needed for reionization could not have formed before $z\sim30$ [\cite{Simulation}]. While these two bounds on the reionization history are well understood, we are yet to uncover the complete history. The observations that shed light on this history can be divided into two classes, direct and indirect probes. As a direct detection, spectroscopy of high redshift quasars or galaxies [\cite{Mortlock2011,Schenker2014}] reveals the absorption by neutral hydrogen, that reveals its fraction as a function of redshift. Spectroscopy of high redshift quasars~[\cite{Becker:2001ee}] shows evidence of IGM transition between its neutral and ionized state around redshift ($z$) 6. 21cm observation [\cite{LoebBook,Loeb,EDGES}], which is beyond the scope of this paper, is the next promising avenue to constrain the epoch of reionization. Observation of luminosity from possible sources of reionization gives an indirect estimation of the reionization history. However, high energetic and fainter sources of reionization are difficult to observe. Until now we have observations of luminosity from sources up to $z\sim10$. We have also been able to estimate the neutral hydrogen fractions till $z\sim8$, albeit with large uncertainties. Scattering of the Cosmic Microwave Background (CMB) photons by the electrons freed during the ionization process provides an alternate window to probe its history. The E-mode auto-correlation spectrum here constrains the integrated free electron fraction along the line of sight. Using these direct and indirect windows of observation, several efforts of constraining reionization history have been made~[\cite{Hu2003,Mitra2013,Douspis2015,Heinrich_2016_PCA,price,Hazra_2017,Paoletti_2019,Hazra_2020,Qin:2020xrg,Chatterjee:2021ygm}] in the last two decades. While comparing with data, the theory of ionization or the free electron fraction is modelled with parametric and non-parametric methods. CMB data releases have been providing crucial information about the line-of-sight integral of free electron fraction. With the Planck 2018 release the most precise information has been obtained about the optical depth with their low multipole E-mode polarization data from the high frequency instrument. In the last decade, in parallel, the detection of sources reached a milestone with Hubble Frontiers Fields (HFF). Here we make a joint analysis of these data in a new model-independent framework. Note that several simpler and popular parametrizations of reionization history in CMB also exist where the free electron fraction is parametrized [\cite{TanhWMAP,Heinrich_2016_PCA,Hazra_2017,Millea}]. However, since these methods do not have source function parametrization, they can not be used with the observation of UV luminosities and in certain cases, they lead to unphysical histories of reionization.    

In this paper, we explore two important aspects of reionization history. Using the model-dependent parametric formalism, we address the UV luminosity density data estimated from the observation of HFF with two different truncation magnitudes. With Gaussian process regression, we make model-independent reconstruction of sources and compare the reconstruction with parametric models. With this model-independent reconstruction of sources, we then reconstruct the reionization history by solving the ionization equations, using CMB and neutral hydrogen data along with the HFF data. The paper is organized as follows -- in the next~\autoref{sec:methods} we discuss the methodology of the parametric model and non-parametric Gaussian process regression. Following that we discuss the datasets used in this analysis in~\autoref{sec:data}. The results are presented in~\autoref{sec:res}. Finally, we conclude in~\autoref{sec:concl}.


\section{Methodology} \label{sec:methods}

The ionization equation describes the time evolution of the volume filling factor of ionized hydrogen in the intergalactic medium, $Q_{H_{II}}$, by a first order ordinary differential equation [\cite{Madau1999,Loeb2003}]:
\begin{equation}
   \frac{dQ_{H_{II}}}{dt}= \frac{\dot{n}_{ion}}{\langle n_H\rangle} -\frac {Q_{H_{II}}}{t_{rec}}
   \label{eq:ioneqn}
\end{equation}
The source term is characterized by the rate of production of ionizing photons $\dot{n}_{ion}$ which depends on the UV luminosity density function $\rho_{UV}$, the efficiency of the source to produce ionizing photons $\xi_{ion}$, and the fraction of photons that actually escape into the IGM $f_{esc}$, and is defined as $\dot n_{ion}=  \rho_{UV}\langle f_{esc}\xi_{ion}\rangle$, where $\langle f_{esc}\xi_{ion}\rangle$ is a magnitude-averaged product.
The sink term in the ionization equation accounts for the recombination process in the IGM. The recombination time $t_{rec}$ is determined by the recombination coefficient $\alpha_B(T)$ and the clumping factor $C_{H_{II}}$; $t_{rec}=\left[ {C_{H_{II}}\alpha_B(T)\left(1+\frac{Y_p}{4X_p}\right)\langle n_H\rangle (1+z)^3} \right]^{-1} $.
We keep the IGM temperature $T$ fixed at 20,000K in our analysis.
$X_p, Y_p$ are the primordial mass fractions of hydrogen and helium respectively.
$n_H, n_{He}, n_{H_{II}}$ are the number densities of hydrogen, helium, ionized hydrogen respectively.

The evolution of the UV luminosity density with redshift can be obtained by parametric and non-parametric methods. In this article, as parametric method we assume the density to be described by a single power-law [\cite{singlepowerlaw1,singlepowerlaw3}] and double power-law [\cite{Ishigaki_2015,Ishigaki_2018}] forms,
\begin{eqnarray}
    \rho_{UV}(z)&=&\rho_{UV_1}\cdot10^{-az}~\label{eq:powerlaw1}\\
    \rho_{UV}(z)&=&\frac{2\rho_{UV,z=z_1}}{10^{a(z-z_1)}+10^{b(z-z_1)}}~\label{eq:powerlaw2}    
\end{eqnarray}

While the first parametrization can be characterized completely by two parameters, namely the tilt ($a$) and the amplitude ($\rho_{UV_1}$), the second needs four parameters: the amplitude ($\rho_{UV,z=z_1}$), two tilts ($a,b$) and the redshift ($z_1$) at which the tilt in the power changes. 

While parametric methods are useful, the functional form restricts their ability to address the data in several instances. Therefore model-independent reconstructions like~\cite{Hazra_2020},~\cite{Mason} provide a conservative analysis owing to their flexibility. In this article, for a Bayesian, non-parametric reconstruction of luminosity density we make use of Gaussian process regression. A Gaussian process (GP) is a collection of random variables such that the joint distribution of any finite subset of it is described by a multivariate Gaussian [\cite{Rasmussen}]. A GP is characterized by its mean function $\mu(\mathrm{x})$ and covariance function  $k(\mathrm{x},\mathrm{x'})$, 
where for a real process $f(\mathrm{x})$, we have $\mu(\mathrm{x}) = \mathbb{E}[f(\mathrm{x})]$, and
$k(\mathrm{x},\mathrm{x'}) = \mathbb{E}[(f(\mathrm{x})-\mu(\mathrm{x}))(f(\mathrm{x'})-\mu(\mathrm{x'}))]$.
For a finite set of training points $\mathrm{x}=\{x_i\}$, a function $f(\mathrm{x})$ evaluated at each $x_i$ can be represented by a random variable with a Gaussian distribution, such that the vector $\mathrm{f}=\{f_i\}$ has a multivariate Gaussian distribution given as $\mathrm{f}\sim \mathcal{N}\left(\mu(\mathrm{x}), C(\mathrm{x},\mathrm{x})\right)$, where $C$ is the covariance matrix characterized by the covariance function $k$, which gives the covariance between two random variables.
For the purpose of our analysis, we have chosen the covariance function to be the Radial Basis Function represented as $ k(x_i,x_j)= \sigma_f\text{exp} \left(- \frac{(x_i-x_j)^2}{2l^2}\right)$, with the correlation length $l$ and amplitude $\sigma_f$.

In our analysis, we define the UV luminosity density in a few redshift nodes, and the values of the UV luminosity density at the these nodes are taken as free parameters. We use the~\texttt{sklearn} [\cite{scikit-learn}] package in \texttt{Python} to implement Gaussian process regression.
We have also developed independent Fortran codes that are used to obtain the bounds on hyperparameters.
Gaussian processes have been useful in the reconstruction of several estimators related to cosmological processes such as Dark Energy [\cite{Holsclaw_2010,Shafieloo,Seikel,Shafieloo_2013}], test of  the standard model [\cite{Aghamousa_2017}] with CMB etc. While we introduce the use of GP in obtaining reionization history by reconstructing the UV luminosity densities, for recombination time, we use the aforementioned analytical formalism as it has been demonstrated in~\cite{Hazra_2020,Mason,Paoletti2021} that apart from strict monotonic histories, $t_{rec}$ can not be constrained due to degeneracies with source terms.

Using the solutions of the ionization~\autoref{eq:ioneqn}, the Thomson scattering optical depth is defined as:
\begin{equation}
    \tau=\int  \frac{c(1+z)^2}{H(z)}Q_{H_{II}}(z) \langle n_{H}\rangle\sigma_T \left(1+\eta\frac{Y_p}{4X_p}\right) dz,
\end{equation}
where helium is assumed to be singly ionized for $z>4$ ($\eta=1$) and doubly ionized for $z<4$ ($\eta=2$) [\cite{Kuhlen}]. 

\section{Data} \label{sec:data}
We have used datasets from observations of 3 different types. Firstly, we use the optical depth constraints from Planck 2018 release of Cosmic Microwave Background observation. With the low multipole polarization likelihood using data from High Frequency Instruments, the best CMB constraint on the optical depth is obtained as $\tau=0.054\pm0.007$~[\cite{Planck2018:param}]. Since it has been demonstrated that the degeneracy between the primordial spectral tilt and the optical depth has reduced significantly in the new CMB data [\cite{Hazra_2020,Paoletti_2019}], instead of using the full data we use this constraint on optical depth as the summary statistics. Secondly we use the derived UV luminosity density data [\cite{Bouwens_2015,Ishigaki_2018}] in the compilation used in~\cite{Ishigaki_2018} analysis, from HFF observations~[\cite{HFF2},~\href{http://www.stsci.edu/hst/campaigns/frontier-fields/}{http://www.stsci.edu/hst/campaigns/frontier-fields/}]. Here we use the luminosity density measurements with truncation magnitude $M_{trunc}$ of -17 and -15. Finally we use the measurements of neutral hydrogen fractions from the Lyman-$\alpha$ emission from galaxies~[\cite{Ono2012,Schenker2014,Tilvi2014,Mason:2019ixe}], damping wings of Gamma Ray Bursts~[\cite{Totani:2005ng,McQuinn:2007gm}] and Quasars spectra~[\cite{Schroeder2013,Greig:2016vpu,Davies:2018pdw}], dark gap in quasar spectra~[\cite{gap}] ionized zones near high redshift quasars~[\cite{Mortlock2011,Bolton2011}].  

\section{Results} \label{sec:res}

\subsection{Can a power-law explain the UV luminosity density data?}
\begin{figure*}
    \centering
       \includegraphics[width=0.495\columnwidth]{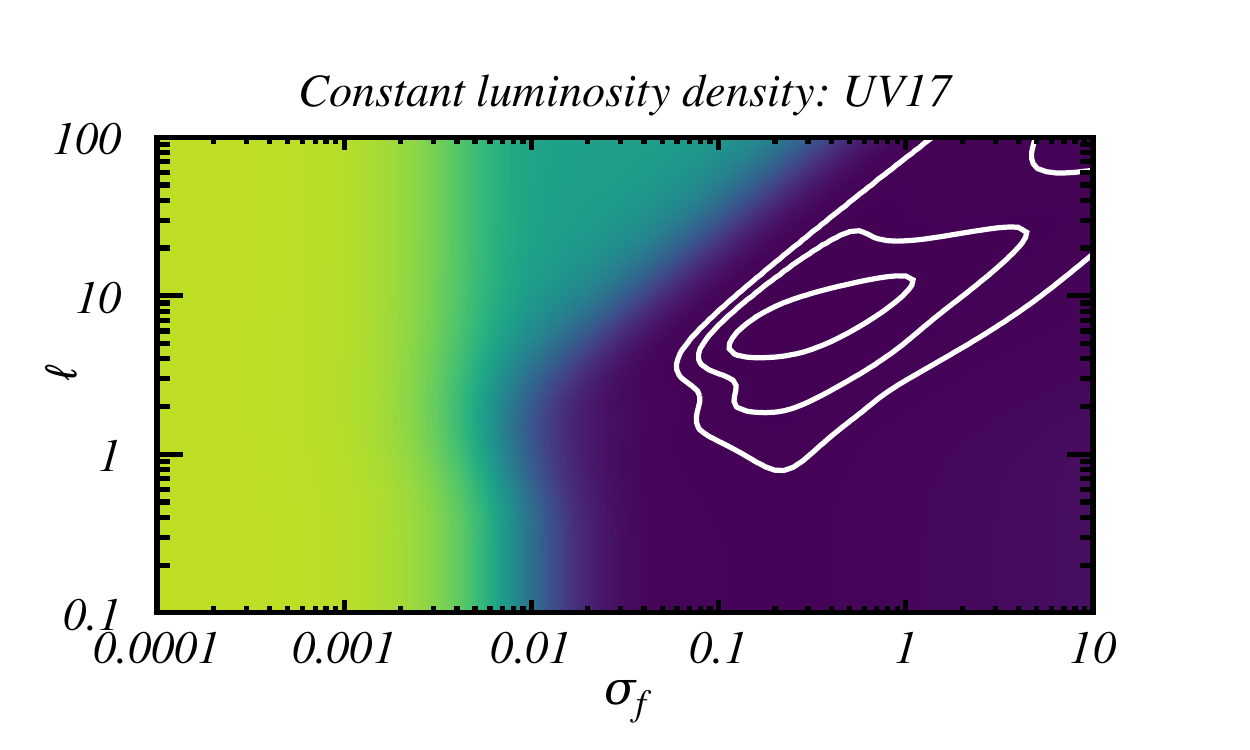}
       \includegraphics[width=0.495\columnwidth]{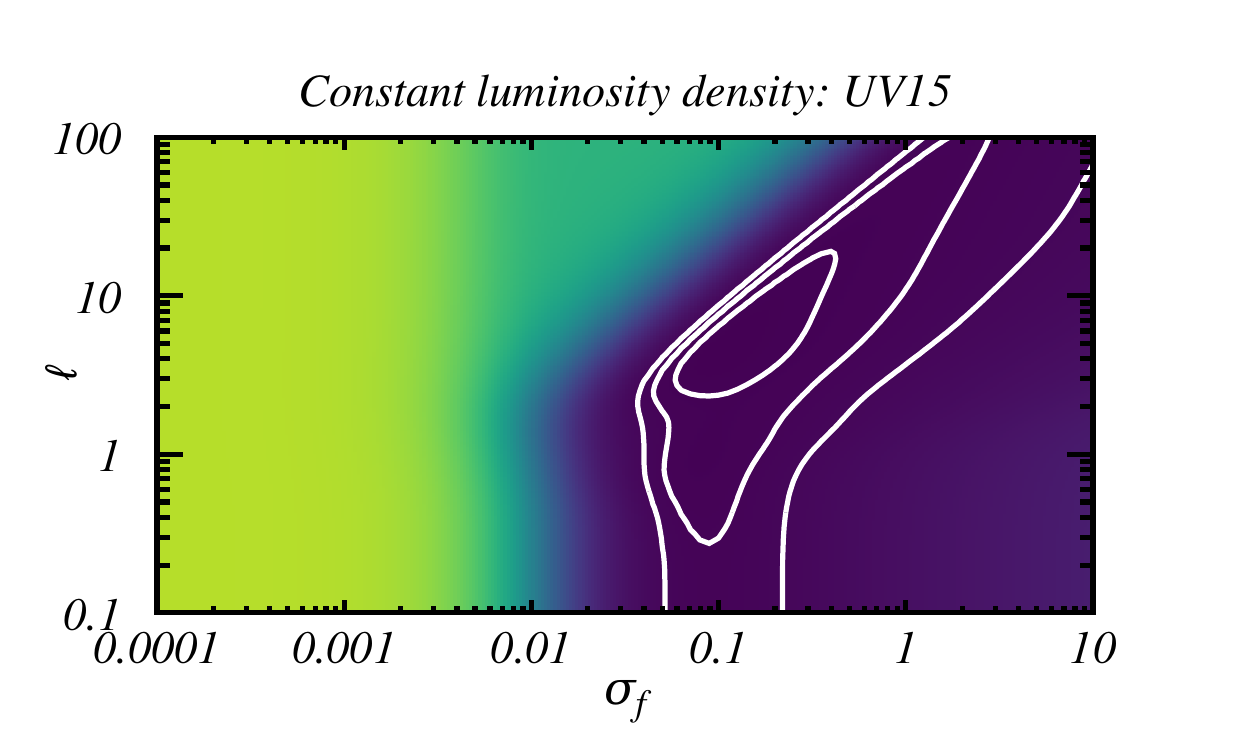}

       \includegraphics[width=0.495\columnwidth]{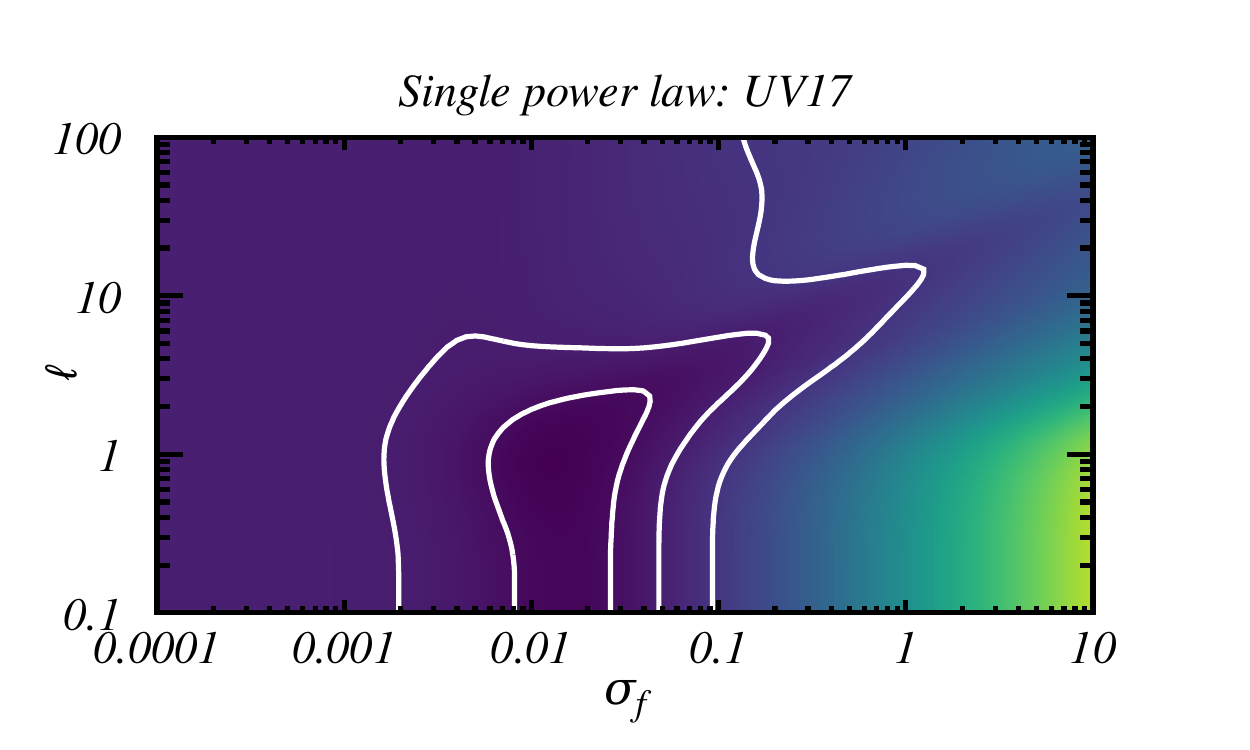}
       \includegraphics[width=0.495\columnwidth]{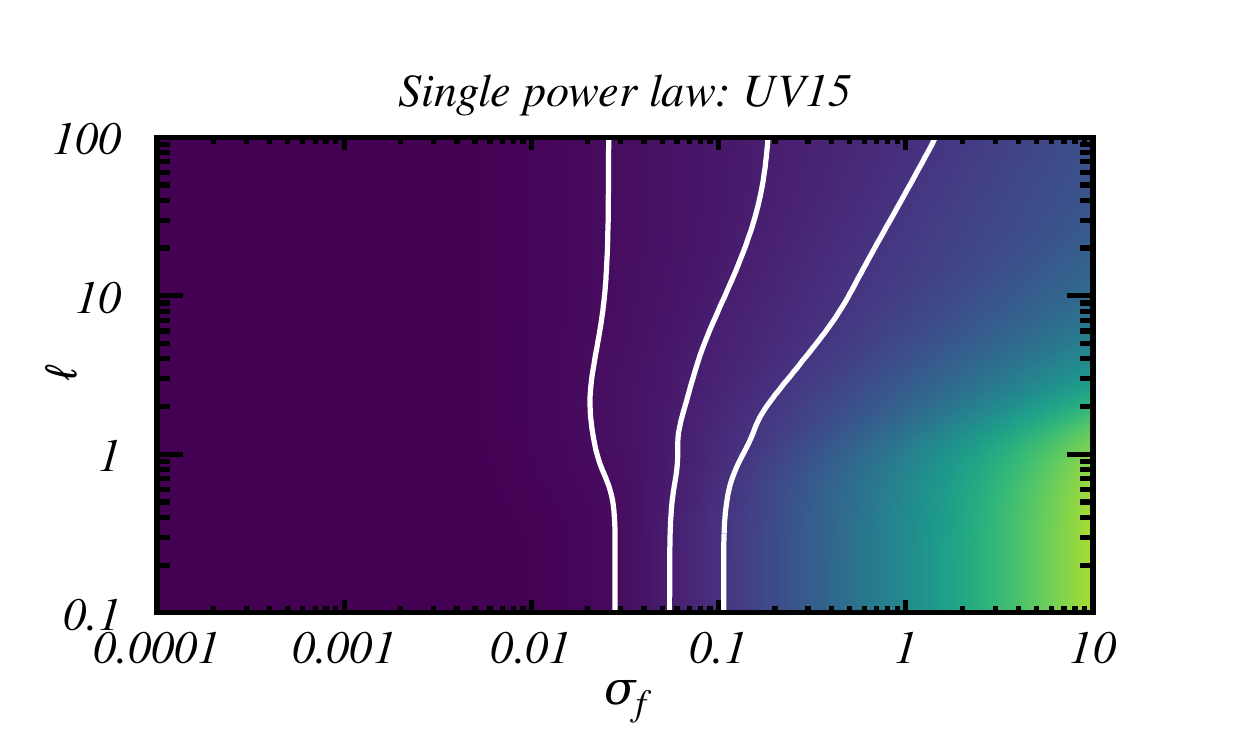}

       \includegraphics[width=0.495\columnwidth]{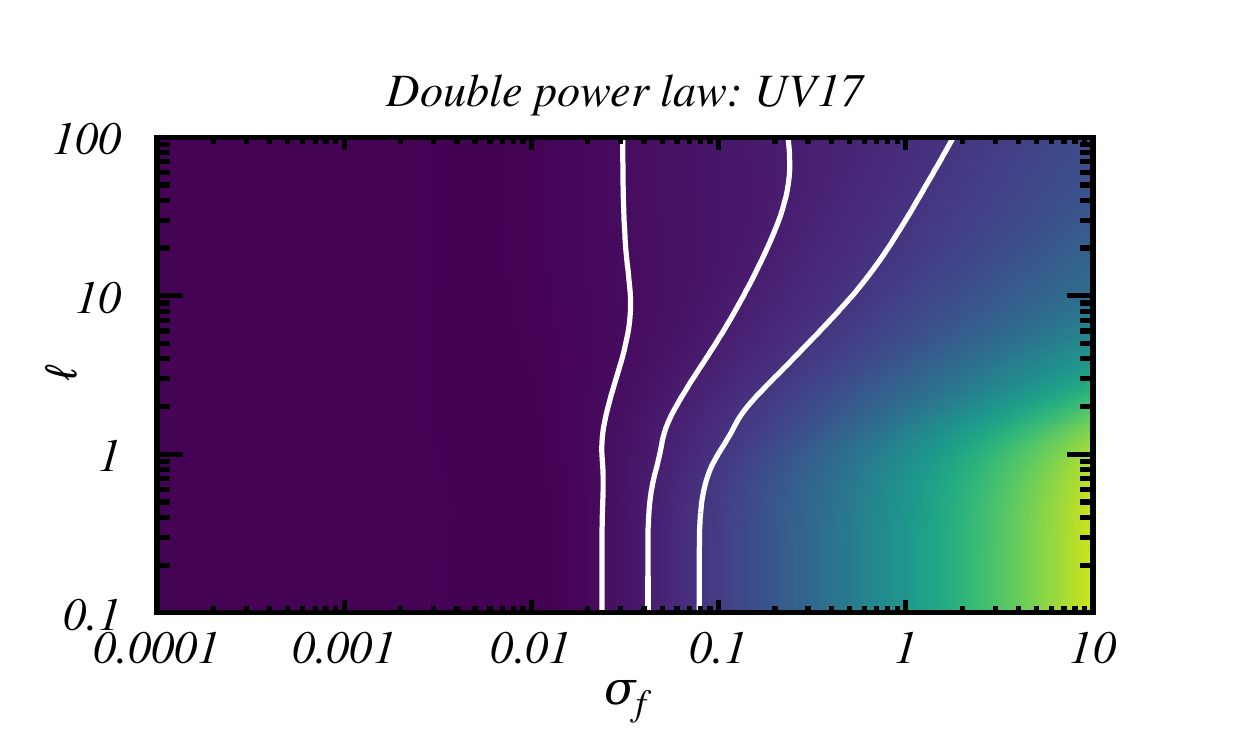}
       \includegraphics[width=0.495\columnwidth]{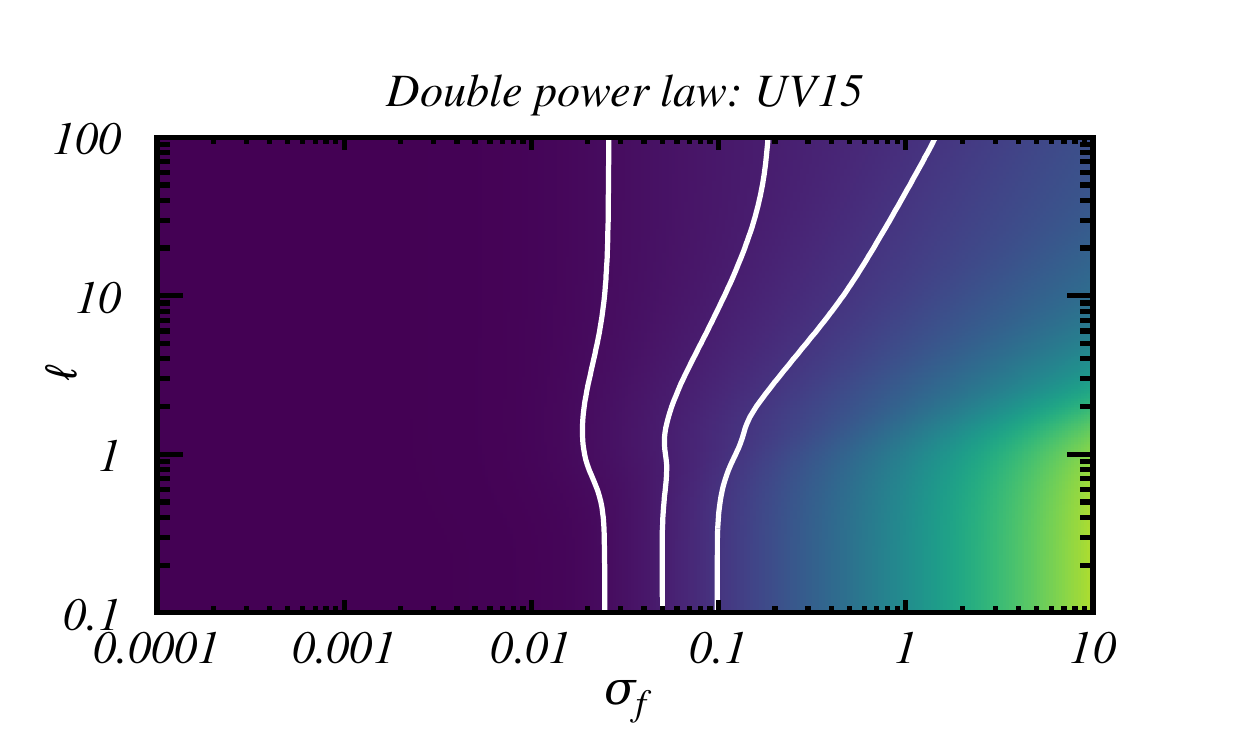}
       
 \caption{~\label{fig0}The log of the GP likelihood, in particular $-2\ln {\cal L}$ are plotted (in color gradient) for different choices of mean functions. They are plotted against the GP hyperparameters $\sigma_f$ and $\ell$ for the choice of Exponentiated quadratic kernel/radial basis function kernel. Left and right columns represent the results from UV17 and UV15 data respectively. Darker regions represent better likelihood. We provide contours upto 3 standard deviations. The $1-3\sigma$ contours can be identified as the lines move towards the brighter colors. The top row contains the results when constant luminosity density over time is assumed to be the mean function. The middle row contains the results for single power law mean function and the bottom row contains the status of logarithmic double power law.}
\end{figure*}

\begin{figure*}
    \centering
       \includegraphics[width=\columnwidth]{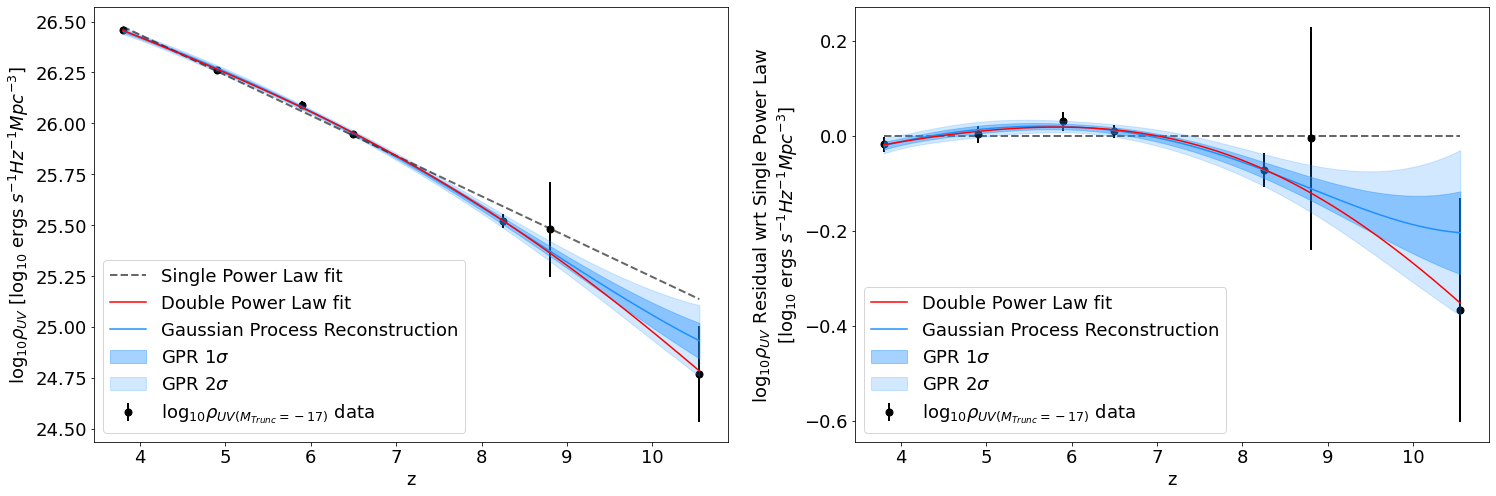}
       \includegraphics[width=\columnwidth]{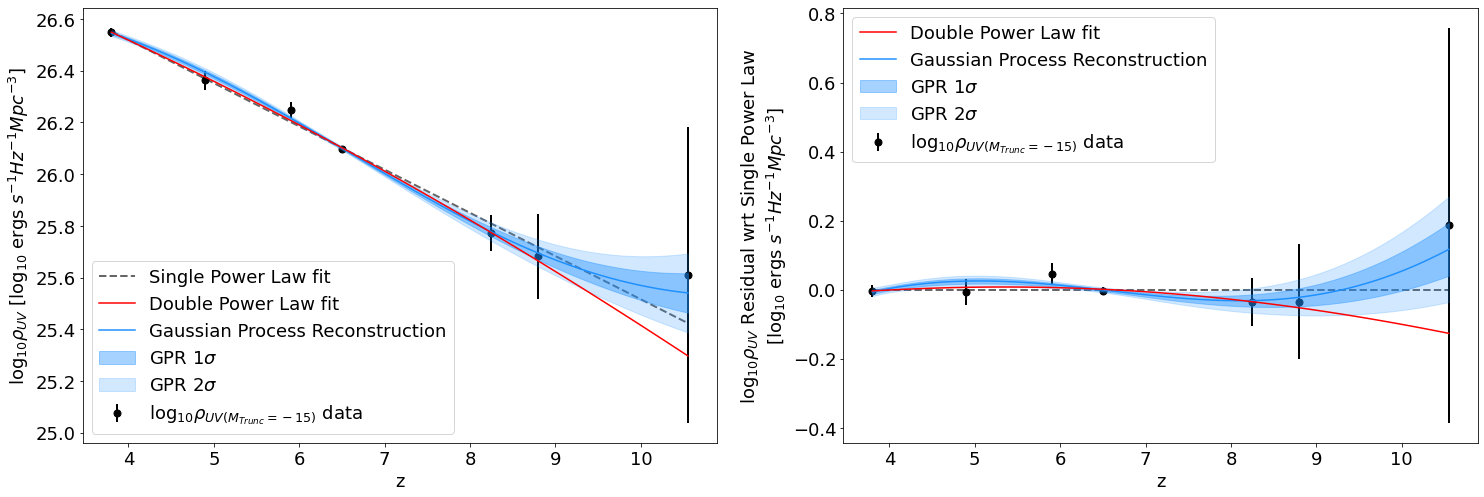}
 \caption{~\label{fig1}Gaussian Process reconstruction of the UV luminosity density using the UV17 (top) and UV15 (bottom) data sets as training points. Plots in the right panel represent same plots in the left but in the residual space {\it w.r.t.} single power-law best fits.}
\end{figure*}

We perform a Gaussian Process Regression (GPR) for the UV luminosity density data for both UV17($M_{trunc}=-17$) and UV15($M_{trunc}=-15$). In~\autoref{fig0} we plot the log of GP likelihood as a function of the hyperparameters. We present the results for three different choices of the mean function: constant UV luminosity density (top), single power law model (middle) and logarithmic double power law model (bottom). The mean functions are chosen as the best-fitted function to the corresponding datasets. 1-3$\sigma$ confidence contours are plotted in each case. If amplitude parameter $\sigma_f$ is well-bounded for a particular mean function, it implies the rejection of that mean function with corresponding significance. As expected, constant mean functions are rejected with high significance. The single power law model is rejected by the UV17 data at more than 2$\sigma$ C.L. The UV15 data however does not rule out the single power law model for higher uncertainties and larger values of UV luminosity density at high redshifts (this does not require drop in power after a particular redshift). The logarithmic double power law model is completely consistent with both the datasets.
We now compare power-law best-fits within the uncertainty bands of this reconstruction as a function of redshifts.~\autoref{fig1} shows the model-independent reconstruction of the UV luminosity density using GPR. The training points here are the $\log_{10}\rho_{UV}$ data from UV15 and UV17, which are used to train the kernel hyperparameters. Then, the optimized hyperparameters are used to interpolate between the training points. The top panel shows the results for UV17 and the bottom panel for UV15. The plots on the left show the interpolation of $\log_{10}\rho_{UV}$ obtained using GPR, overlaid with the corresponding data points used as training points, along with the power-law best fits. 

We perform $\chi^2$ minimization to obtain the best-fit parameters for the two power-law (single and logarithmic double power-law) forms. The plots on the right show the residuals of power-law, GPR construction, and data points with respect to the single power-law. 

Therefore, the zero line represents the single power-law. In the top panel, for UV17, we notice that for higher redshifts the single power-law does not fall within 95\% confidence interval of the model-independent reconstruction and therefore it is ruled out (as expected, this is consistent with the results from the hyperparameter contours in~\autoref{fig0}). The double power-law seems to be agreeing with the reconstruction at all redshifts in this case. In case of UV15 (bottom panel), it can be observed that power-law models agree with the model-independent reconstruction with small deviations. In our analysis with constant, single power law and logarithmic double power law as mean functions, we find that the double power law is completely consistent with both UV15 and UV17 data. While single power law is consistent with UV15 data, it is ruled out at more than 95\% C.L. by UV17 data.

\subsection{Constraints on reionization history}

To solve the ionization equation (\autoref{eq:ioneqn}), instead of assuming a specific function for the UV luminosity density, we follow a model-independent approach. Compared to earlier model-independent approaches in~\cite{Hazra_2020,Mason}, we use Gaussian Process Regression to construct the UV luminosity density at different redshifts and find constraints on reionization history by solving the ionization equation using this model-independent form.
 
We follow similar approach to~\cite{Gerardi:2019obr}. Here we use 4 equidistant nodes between redshifts 4-10 to define the UV luminosity densities. The redshift range is chosen in such a way that it falls completely inside the range of redshifts of available data. It should be noted that to capture this correlation the distances between the nodes have to be smaller than the correlation length and we have tested the robustness of the analysis with varying number of nodes. The values of UV luminosity density at the redshift nodes are taken as free parameters for Markov Chain Monte Carlo (MCMC) sampling (using {\tt emcee} [\cite{emcee}]), and at each step these points are used as training points for Gaussian Process Regression. Therefore, here the Gaussian process reconstruction provides the samples of the history of UV luminosity densities for different choices of training points. Using the source function obtained from the sample of UV luminosity density evolution, we then solve~\autoref{eq:ioneqn} to get the reionization history. For the integration we need $\rho_{UV}(z)$ to extend beyond the available data. Since outside the range of data, GP reconstruction simply merges with the mean function, the choice of mean function is important here. We have used logarithmic double power law as the mean function for the GP here as both datasets agree with this functional form. While for UV15, single power law would also act as an appropriate choice of mean function, to keep the uniformity, here too we use double power law mean function that includes single power law within its parameters.
\begin{figure*}
    \centering
    \includegraphics[width=0.6\columnwidth]{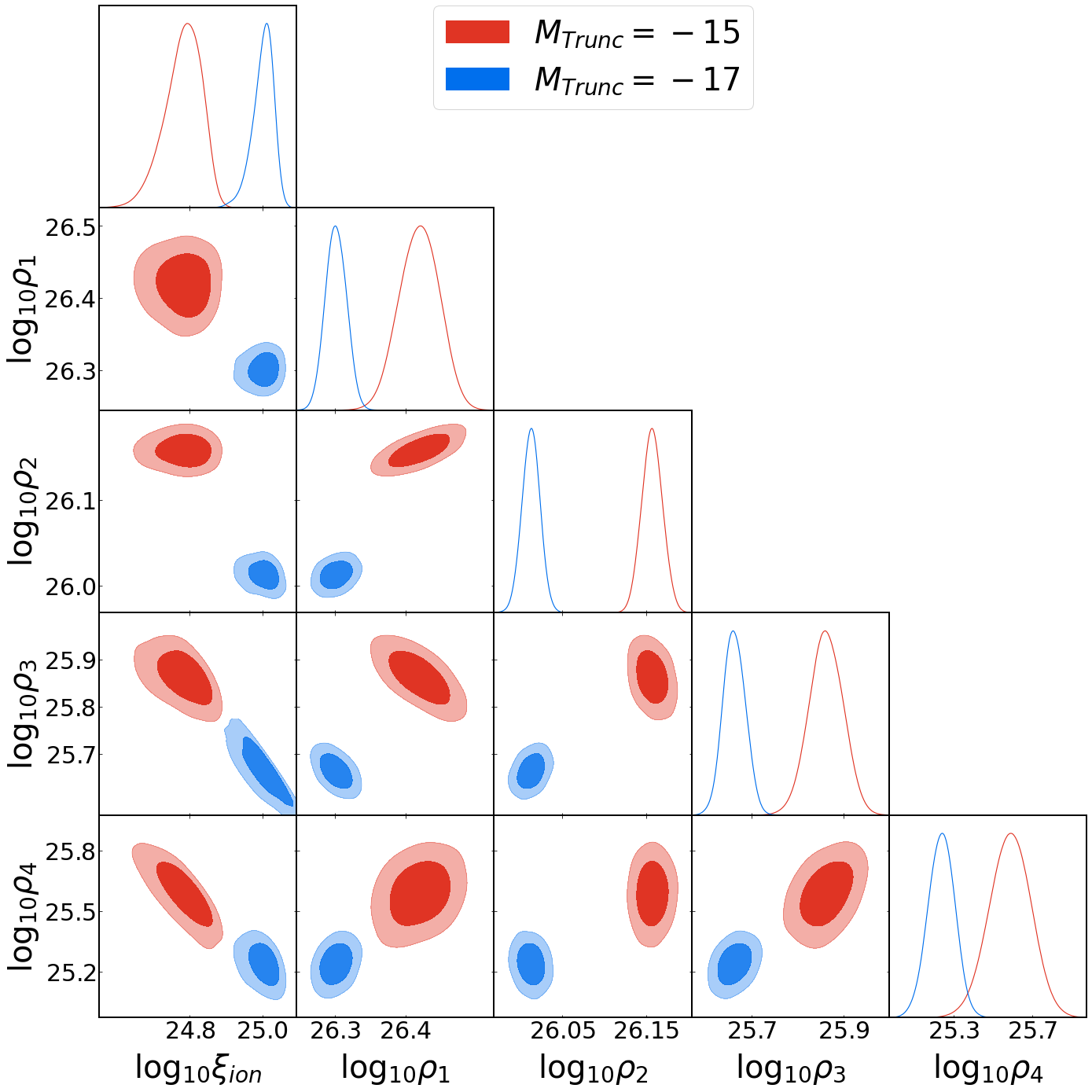}
        \includegraphics[width=0.39\columnwidth]{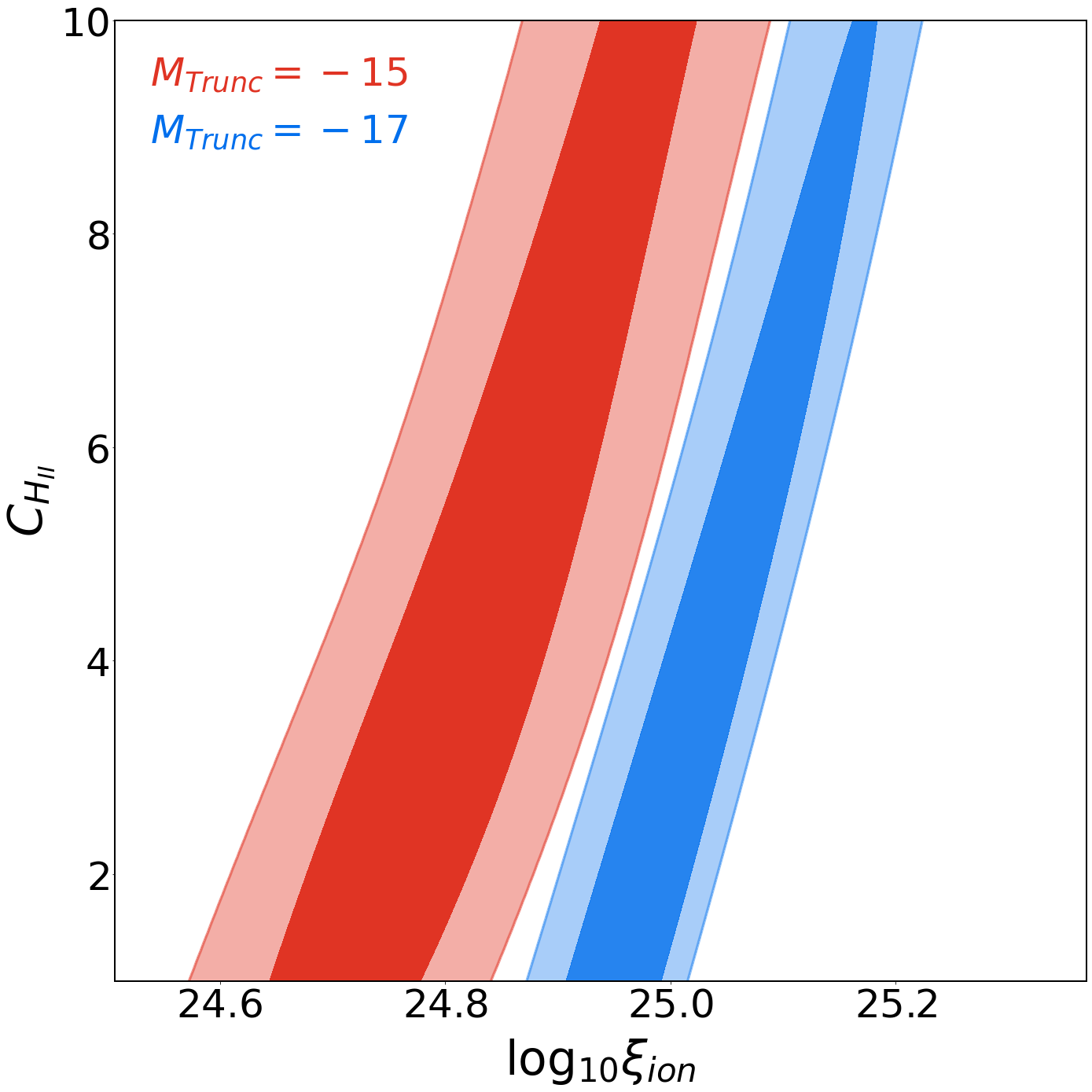}
    \caption{~\label{fig3} [Left] Posterior distributions of 5 free parameters of the analysis with the UV luminosity density constructed using Gaussian process regression for the values $\rho_{1-4}$. Note that here $C_{H_{II}}$ is kept fixed. [Right] 1-2$\sigma$ confidence contours for $C_{H_{II}}$ and $\log_{10}\xi_{ion}$ [in $\log_{10}$ Hz erg$^{-1}$] obtained from our model-independent analysis using combination of \textit{Planck} optical depth, $Q_{H_{II}}$ and UV luminosity density data for $M_{trunc}=-15$ and $-17$. Note that the clumping factor is essentially degenerate with the efficiency factor as shown in~\cite{Gorce,Mason}}
\end{figure*}

We now obtain joint constraints on reionization history using all the 3 data sets described earlier: \textit{Planck} optical depth, $Q_{H_{II}}$ data and UV luminosity density data for $M_{trunc}=-17$ and $-15$. In the ionization equation, we vary $\langle f_{esc}\xi_{ion}\rangle$ as one parameter by absorbing $f_{esc}$ into the parameter $\xi_{ion}$ (see review~\cite{Dayal:2018hft} for bound on the escape fraction). We keep a uniform prior on $\log_{10}\xi_{ion}$ between 23.5 and 27.5 [$\log_{10}$ Hz erg$^{-1}$] as suggested in~\cite{price}. We perform analyses with the clumping factor fixed as well as free with a uniform prior between 1 and 10. The other free parameters in the analysis are the values of UV luminosity density at the four redshift nodes: $\rho_{1-4}$. For the case with fixed $C_{H_{II}}$, left panel of~\autoref{fig3} shows the MCMC constraints obtained for the 5 free parameters using both the combinations of data sets: CMB+QHII+UV17 (in blue) and CMB+QHII+UV15 (in red). The constraints on UV luminosity densities at different redshift nodes are significantly different in these two data combinations, and since UV15 integrates luminosity function till larger magnitudes, we find constraints on the density to be shifted at larger values consistently at all redshift nodes. At higher redshifts, the constraints degrade as the data has larger uncertainties. Since the source term depends on both the UV luminosity density and the ionizing photon production efficiency, they must be anti-correlated in the CMB+UV+QHII data combination. This is easy to appreciate, as higher UV luminosity density and higher efficiency will lead to a larger optical depth which will be ruled out by CMB optical depth constraints. We notice this anti-correlation in the last two redshift nodes as they mainly fall {\it within} the epoch of reionization. As expected, we also notice that the marginalized posteriors of $\log_{10}\xi_{ion}$ are shifted in opposite directions compared to the UV luminosity density posterior distributions. 

\begin{figure}
    \centering
    \includegraphics[width=\columnwidth]{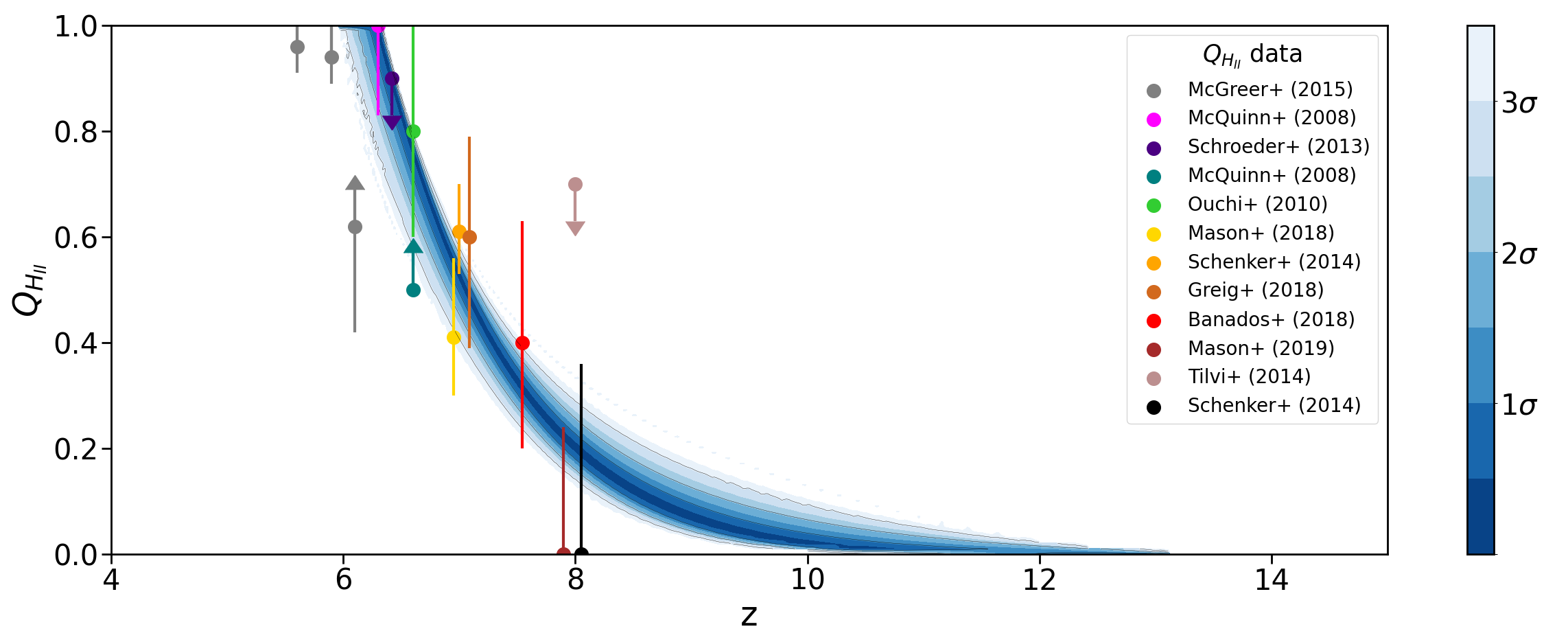}    
    \includegraphics[width=\columnwidth]{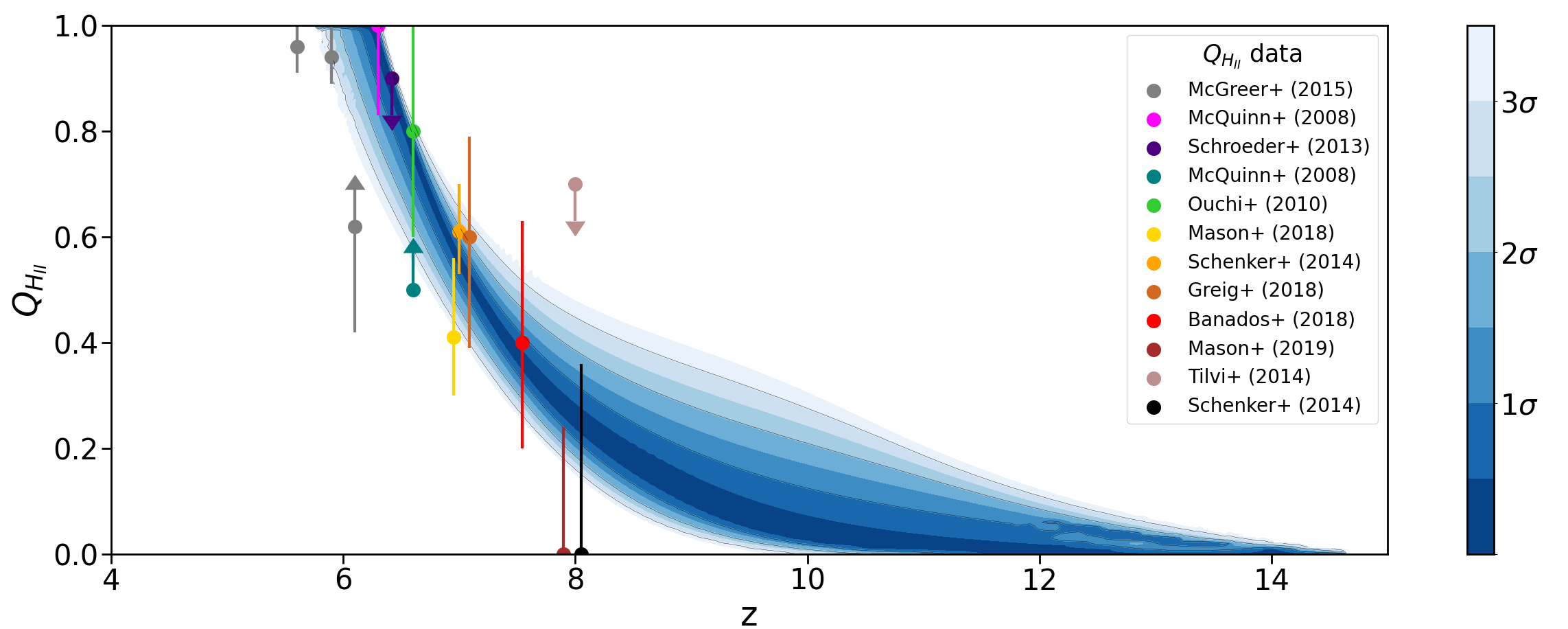}    
    \caption{~\label{fig4}The reconstructed history of reionization obtained from the MCMC samples when CMB optical depth, UV luminosity density data (top: UV17, bottom: UV15) and QHII data (plotted with references) are used (corresponding to~\autoref{fig3}).}
\end{figure}

The resulting constraints on the redshift of reionization $z_{re}$, duration of reionization $\Delta z$ and the reionization optical depth $\tau$ are listed in \autoref{table1}. The optical depth constraints obtained in both the cases agree with 1$\sigma$ optical depth from \textit{Planck} results. We find that the optical depth is larger in the analysis where UV15 is used compared to UV17 data. Our constraint on the duration of reionization ($\Delta z$) indicates that the significant part of the reionization (10\% to 90\% ionization) has occurred in $\sim 2$ (for CMB+UV17+QHII) and $\sim 3$ redshifts (for CMB+UV15+QHII). Since UV15 data shows flattening of the luminosity densities at higher redshifts, it allows for a longer and earlier history of reionization compared to the data combination where UV17 data is used. The 68\% and 95\% bounds indicate that the marginalized posterior of $\Delta z$ is significantly skewed. The redshift where the reionization is 50\% completed is denoted as the redshift of reionization ($z_{re}$) and for both data combinations we find $z_{re}\sim7$.

For the case with $C_{H_{II}}$ assumed as a free parameter, we plot the correlation between the $\log_{10}\xi_{ion}$ and $C_{H_{II}}$ in the right panel of~\autoref{fig3}. We find that the efficiency and the clumping factors are positively correlated and both can not be constrained with these data combinations. This degeneracy is also discussed in~\cite{Gorce,Mason,Paoletti2021}. This indicates anti-correlation with the recombination time. A decrease in the ionizing photon production efficiency decreases the source term which is only compensated by increased time for recombination to sustain ionization.

\renewcommand{\arraystretch}{1.5}
\begin{table}[h]
    \centering
    \begin{tabular}{|c|c|c|}
    \hline
         & CMB+UV17+QHII 
         & CMB+UV15+QHII 
         \\
         \hline
         $\tau$ 
         & $0.052\pm0.001\pm0.002$
         & $0.056_{-0.004-0.005}^{+0.002+0.006}$
         \\
         \hline
         $\Delta z$
         & $2.05_{-0.21-0.30}^{+0.11+0.37}$
         & $2.98_{-1.0-1.1}^{+0.30+1.8}$ 
         \\
         \hline
         $z_{re}$ 
         & $6.93_{-0.04-0.13}^{+0.08+0.11}$
         & $7.00_{-0.10-0.22}^{+0.11+0.22}$ 
         \\
         \hline
    \end{tabular}
    \caption{~\label{table1} Joint constraints (with $C_{H_{II}}$ fixed) on the reionization optical depth $\tau$, duration of reionization $\Delta z$ and the redshift of reionization $z_{re}$ with 68\% and 95\% intervals.}
\end{table}

From the samples obtained in CMB+UV17+QHII and CMB+UV15+QHII data analysis (with fixed $C_{H_{II}}$), the reconstructed histories of reionization are plotted in the top and bottom panels respectively of~\autoref{fig4}. We use {\tt fgivenx} [\cite{fgivenx}] package to compute the confidence band on the reconstructed ionization history. Different confidence limits are indicated in the color bar at the right. On top of the band, we have plotted the QHII constraints we have used in our analysis. Note that when UV17 data is used in the combination, the Gaussian process reconstruction does not allow early onset of ionization as it constrains the $Q_{H_{II}}<0.1$ at $z\sim10$ with 95\% C.L. When UV15 data is used we find the constraints are much more relaxed, with $Q_{H_{II}}<0.3$ at the same redshift. Here we also find a longer tail of reionization between $z=8-15$. It is important to note that with the present observation at high redshifts the constraints can not be made tighter unless we restrict our model to a specific redshift symmetric form.

\section{Conclusions} \label{sec:concl}

In this paper, using Gaussian process regression as a non-parametric method for model-independent reconstruction of the evolution of UV luminosity density we address two questions. Firstly, within the confidence band of reconstruction, we show that while the commonly used logarithmic double power-law model of UV luminosity density evolution agrees well with the data, the single power-law is ruled out. Secondly, using the reconstructed UV luminosity density evolution we reconstruct the reionization history using the optical depth from the Cosmic Microwave Background observation, UV luminosity data from Hubble Frontiers Field observation, and neutral hydrogen fraction data from galaxy, quasar and gamma ray burst observations. Gaussian process regression allows a free form reconstruction of reionization history and thereby provides the platform for a conservative analysis. The use of three types of datasets, even in this conservative framework, breaks degeneracies between parameters and provides stringent constraints. From the joint constraints (when UV luminosity data with truncation magnitude 17 is used in the combination) we highlight that at 68\%-95\% C.L. we find the optical depth $\tau=0.052\pm0.001\pm0.002$ and the duration between 10-90\% ionization is $2.05_{-0.21-0.30}^{+0.11+0.37}$ in redshifts. When the joint analysis uses UV luminosity data assuming truncation magnitude of -15, a longer tail in the ionization history is observed between redshift 10-14 with less than 30\% ionization fraction (at 95\% C.L.) at redshifts higher than 10. 

Since we find that the use of UV luminosity density obtained by assuming higher (-15) truncation magnitude results in higher optical depth and larger duration of ionization, high redshift observations with JWST [\cite{JWST,Madau2017}] and THESEUS [\cite{THESEUS}] will definitely be helpful in providing better constraints, particularly around the tail. Cosmic variance limited CMB polarization data [\cite{LiteBIRD}] will also help in constraining the optical depth better and therefore can also indirectly improve the joint constraints.

\begin{acknowledgements}
We would like to acknowledge the use of the HPC system Nandadevi (\url{hpc.imsc.res.in}) at the Institute of Mathematical Sciences, Chennai. We would like to thank Abraham Loeb, Sourav Mitra, Daniela Paoletti, Tirthankar Roy Choudhury and Arman Shafieloo for their comments on the manuscript. DKH thanks Masami Ouchi and Masafumi Ishigaki for providing their UV luminosity density data compilation.
\end{acknowledgements}

\bibliographystyle{aasjournal}
\bibliography{GaussReion}



\end{document}